
\tolerance 2000
\font\titolo=cmbx10 scaled\magstep0

\magnification=\magstep1
\font\myit=cmti7 scaled\magstep2

\parskip 3truemm
\hsize=140 truemm
\vsize=210 truemm
\hoffset=15 truemm

\def\mn{\medskip \noindent}

\def\bn{\bigskip \noindent}

\def\nostrocostrutto#1\over#2{\mathrel{\mathop{\kern 0pt \rlap
  {\raise.2ex\hbox{$#1$}}}
  \lower.9ex\hbox{\kern-.190em $#2$}}}
\def\lsim{\nostrocostrutto < \over \sim}

\def\noi1{noi1}

\def\n{neutralino}

\def\c{\chi}
\def\den{\Omega_{\chi} h^2}
\def\m{m_{\chi}}

\def\tb{\tan \beta}
\def\mh{m_h}
\def\msf{m_{\tilde f}}
\def\eee{E_{\rm ee}}
\def\kev{{\rm KeV}}
\def\gev{{\rm GeV}}
\def\tev{{\rm TeV}}
\def\KeV{{\rm KeV}}
\def\GeV{{\rm GeV}}
\def\TeV{{\rm TeV}}

\nopagenumbers

\bn

\rightline{DFTT 38/93}
\rightline{August 1993}
\bn
\bn
\centerline{\titolo On the Neutralino as Dark Matter Candidate.}
\centerline{\titolo II. Direct Detection.}
\bn
\mn
\centerline{A. Bottino, V. de Alfaro, N. Fornengo, G. Mignola, S. Scopel}
\bn
\mn
\centerline{\myit {Dipartimento di Fisica Teorica
dell'Universit\`a di Torino}}
\centerline{\myit {and INFN, Sezione di Torino, Italy}}
\centerline{\myit {Via P. Giuria 1, 10125 Torino, Italy}}
\bn
\bn
\bn
\centerline{\bf Abstract}
\medskip
\noindent
Evaluations of the event rates relevant to direct search for dark
matter neutralino are presented for a wide range of neutralino masses
and for various detector materials of preeminent interest. Differential
and total rates are appropriately weighted over the local neutralino
density expected on theoretical grounds.

\vfill
\eject

\footline{\hfill}
\headline{\hfil-- \folio\ --\hfil}
\pageno=1

{\bf 1. Introduction}

The most natural way of searching for neutralino dark matter is
provided by direct experiments where the effects induced in
appropriate detectors by neutralino--nucleus elastic scattering may be
measured.  The aim of this paper is to present the evaluation of the
event rates relevant to these direct measurements and to discuss the
experimental sensitivities required in this search for it to be
successful.

Our present analysis refers to a wide range for the neutralino masses,
${\rm 20~GeV} \leq \m \leq {\rm 1~TeV}$,
and examines various detector materials of present
experimental interest: $Ge$, $NaI$, $Xe$, $CaF_2$.
The theoretical framework
is provided by the Minimal
SuSy Standard Model (MSSM), implemented with a general GUT assumption,
as discussed in the companion paper [1] (this reference will
hereafter be referred to as paper I).
The model depends on a number of free parameters which, apart from
the masses of a few particles which come into play, are chosen here
to be $M_2$, $\mu$, $\tan \beta$.  The definition of these parameters is
recalled in paper I.

In the following sections the differential and the integrated rates will be
presented and discussed also in connection with the properties of the
neutralino relic density.

\bigskip
{\bf 2. Event Rates}

   For the nuclear recoil spectrum we use the expression

$${{dR}\over{dE_R}}=
\sum_i{{R_{0,i}}\over{<E_R^{max}>}}F_i^2(E_R)I(E_R)\eqno(2.1)$$

 where

$$R_{0,i}=N_T{{\rho_{\c}}\over{m_\c}}\sigma_i<v>\eqno(2.2)$$

 The various quantities in Eq.s (2.1)--(2.2) are defined as follows:
$E_R$ is the nuclear recoil energy, i.e.
$E_R={{m_{red}^2}}v^2(1-\cos \theta^*)/{m_N}$ ($\theta^*$ is the scattering
angle defined in the neutralino--nucleus center of mass frame,
$m_N$ is the nucleus mass, $m_{\rm red}$ is the neutralino--nucleus
reduced mass and $v$ is the relative velocity),
$E_R^{max}={{2m_{red}^2}}v^2/{m_N}$,
$F(E_R)$ denotes the nuclear form factor,
$\sigma$ is the neutralino--nucleus cross section.
The subscript $i$ refers to the two cases of coherent and spin--dependent
effective interactions.
$N_T$ is the number density of the target nuclei and $\rho_\chi$ is the local
neutralino matter density.
Finally, $I(E_R)$ is given by

$$I(E_R)={{<v^2>}\over{<v>}}
\int_{v_{min}(E_R)}^{v_{max}}  dv{{f(v)}\over{v}}\;.\eqno(2.3)$$

 In Eq.(2.3) $f(v)$ is the velocity distribution of neutralinos in the
Galaxy, measured in the Earth rest frame,
$v_{\rm min}(E_R)$ is given by
$v_{\rm min}(E_R)=({{m_N E_R}/({2m_{\rm red}^2})})^{1/2}$.
The averages
appearing in Eq.(2.2)--(2.3) denote averages over the velocity distribution in
the Earth rest frame. An explicit formula for $I(E_R)$ in the case of a
maxwellian velocity distribution may be found in Ref.[2].

The differential rates that will be presented in Sect.6 will be
expressed in terms of the electron--equivalent energy $E_{ee}$ rather than
in terms of $E_R$. These two variables are simply
proportional: $E_{ee}=Q E_R$, where $Q$ is defined as quenching factor.

In sect.6 we will also discuss the total rates, obtained by integrating
$dR/dE_{ee}$ over appropriate ranges of $E_{ee}$ above the experimental energy
threshold.

\bigskip
{\bf 3. Neutralino--Nucleus Elastic Cross Sections}

The total cross sections for neutralino--nucleus elastic scattering
have been evaluated following the procedure discussed in Ref.s[2--3].

Here we only summarize some of the main properties. Neutralino--quark
scattering is described by the amplitudes with
Higgs boson exchanges and $Z$ boson exchange
in the t--channel
and by the amplitudes with
squark exchanges in the s-- and u--channels.
The neutral Higgs bosons considered here are the two
CP--even bosons: $h,H$ (of masses $m_h$, $m_H$ with $m_H>m_h$) and the CP--odd
one: $A$ (of mass $m_A$).

 The relevant properties for these amplitudes are: 1) Higgs boson
exchanges contribute a coherent cross section which is only vanishing
when there is no zino--higgsino mixture in the neutralino composition[4];
2) $Z$ boson exchange
provides a spin--dependent cross section which takes
contribution only from the higgsino components of $\chi$; 3) squark
exchanges contribute a coherent cross section (due to zino--higgsino
mixing) as well as a spin--dependent cross section (due mainly to the
gaugino components of $\chi$)[5].

In the evaluations of the cross sections presented here, we have adopted for
the masses of the Higgs bosons and of the sfermions the various sets
of values indicated in paper I.
In particular we have relaxed here the assumption,
used in previous papers [2--3],
that $m_{\tilde f} > {\rm 150~GeV}$,
since this stringent lower bound, previously reported
by the CDF Collaboration [6], is significantly weakened by more
refined analyses [7,8]. Thus the following two cases have been considered for
$\msf$:
1) $m_{\tilde f} = 1.2~ m_\c$,  when $m_\c > 45$ GeV,
$m_{\tilde f} = 45$ GeV
otherwise, except for the mass
of the top scalar partners (the only one relevant to radiative corrections)
which has been taken $\tilde m = 3$ TeV
(hereafter this choice for $\msf$ will simply be referred to as:
 $\msf =1.2~\m$);
\ 2) $m_{\tilde f} = 3$ TeV
for all sfermions.
The Higgs boson mass $m_h$ (taken here as a free parameter) is set equal to
values close to its experimental lower bound $\sim 50$ GeV [9].

Let us recall that, whereas the coherent cross section is
proportional to the square of the mass number of the nucleus, the
non--coherent cross section brings a spin--dependence that is commonly
represented by a factor $\lambda^2 J(J+1)$. The values used here
for this last
quantity are taken from [10];
other evaluations may be found in Ref. [11].

   It is worth noticing that, in the model we are considering here, the
event rates for neutralino detection are largely dominated by coherent
effects in most regions of the parameter space. In the small domains
where spin--dependent effects dominate over the coherent ones the total
rates are usually too low to allow detection. Then in the present paper
we do not consider explicitly materials enriched in high--spin isotopes,
even if these materials are very interesting for a search of
hypothetical dark matter particles which interact with matter via substantial
spin--dependent interactions. In particular for $Ge$ and $Xe$ we consider
only natural
isotopic compositions (enrichment in some isotope would not appreciably
modify our results).

\bigskip
{\bf 4. Nuclear Form Factors}

Let us turn now to the $E_R$--dependence introduced in the nuclear recoil
spectrum by the form factors.
These form factors depend sensitively on the nature of the effective
interaction
involved in the neutralino--nucleus scattering.
For the coherent case, we simply employ the standard parametrization[12]

$$F(E_R)={3j_1(qr_0) \over {qr_0}} e^{-{1 \over 2} s^2 q^2}~~. \eqno(4.1)$$

\noindent
where $q^2 \equiv\mid{\bf{q}}\mid^2=2m_NE_R$ is the squared
three--momentum transfer, $s \simeq 1~ \rm fm$ is the
thickness parameter for
the nucleus surface, $r_0 = (r^2-5s^2)^{1/2}$, $r=1.2~A^{1/3}$ fm and
$j_1(qr_0)$ is the spherical Bessel function of index 1.

The form factor of Eq.(4.1) introduces substantial suppression in the recoil
spectrum unless $q r_0 << 1$. A noticeable reduction in $dR/dE_R$ may already
occur at threshold $E_R=E_R^{\rm th}=\eee^{\rm th}/Q$ when
$r_0 \sqrt{2 m_N E_R^{\rm th}}$ is not small compared to unity. The actual
occurrence of this feature depends on a few parameters of the detector
material:
nuclear radius, quenching factor, threshold energy $\eee^{\rm th}$. The values
of these parameters for the nuclei considered in this
paper are reported in Table 1 [13].
Also the values of $F^2(\eee^{\rm th})$ calculated from Eq.(4.1) are
given in this same table. We see that the effect of the form factor
is very moderate for $^{19}F$,$^{23}Na$ and $^{76}Ge$. In $^{76}Ge$,
as compared to $^{19}F$ and $^{23}Na$, this occurs
since the larger nuclear extension is compensated by a smaller value of
$\eee^{\rm th}$ and a larger $Q$. The effect of the form factor on the shape of
the differential spectrum as a function of $\eee$ depends mainly on how close
to
the threshold energy $\eee^{\rm th}$ the first zero of the function
$j_1(q r_0)/(q r_0)$ occurs (this zero is located at
$\eee^0 \simeq  10 Q / (m_N r_0^2)$). Comparing the values of
$\eee^{\rm th}$ and of $\eee^0$ reported in Table 1, we see that only for
$^{127}I$ (and marginally for $^{131}Xe$) this effect is expected to be
noticeable.

In general for the spin--dependent case there are no
analytic expressions for the form factors.
However, numerical analyses have been performed on
$^{131}Xe$ [12], $^{93}Nb$ [18] and $^{73}Ge$ [19].
The general feature is that these form factors
have a much milder dependence on $E_R$ as compared to the coherent ones,
because only a few nucleons participate in the neutralino--nucleus scattering
in
this case.
In our evaluations we use the results of Ref.s [12,19] for
$^{131}Xe$ and $^{73}Ge$.
For $^{127}I$ we assume that the form factor has the same behaviour as
in the case of $^{131}Xe$.
For $^{23}Na$ and $^{19}F$ we employ the analytic expression (4.1),
since for these light nuclei the suppression due to
nuclear size is small anyway.

\bigskip
\centerline{\bf Table 1}
$$
\vbox{ \offinterlineskip \hrule
\halign{ \vrule#&
  \strut\quad \hfil#\hfil\quad&&\vrule#&
  \strut\quad \hfil#\hfil\quad\cr
&\omit&&\omit&&\omit&&\omit&&\omit&\cr

&Nucleus\hfil&&$Q$&&$E_{\rm ee}^{\rm th}$ (KeV)&&
$F^2(E_R^{\rm th})$&&$\eee^0 (\rm KeV)$& \cr

&\omit&&\omit&&\omit&&\omit&&\omit&\cr
\noalign{\hrule}
&\omit&&\omit&&\omit&&\omit&&\omit&\cr

&$^{19}F$&&0.09&&8.5&&0.84&&378&\cr

&\omit&&\omit&&\omit&&\omit&&\omit&\cr
\noalign{\hrule}
&\omit&&\omit&&\omit&&\omit&&\omit&\cr

&$^{23}Na$&&0.23&&4&&0.96&&630&\cr

&\omit&&\omit&&\omit&&\omit&&\omit&\cr
\noalign{\hrule}
&\omit&&\omit&&\omit&&\omit&&\omit&\cr

&$^{40}Ca$&&0.07&&8.5&&0.44&&62&\cr

&\omit&&\omit&&\omit&&\omit&&\omit&\cr
\noalign{\hrule}
&\omit&&\omit&&\omit&&\omit&&\omit&\cr

&$^{76}Ge$&&0.25&&2&&0.86&&61&\cr

&\omit&&\omit&&\omit&&\omit&&\omit&\cr
\noalign{\hrule}
&\omit&&\omit&&\omit&&\omit&&\omit&\cr

&$^{127}I$&&0.07&&4&&0.05&&7.4&\cr

&\omit&&\omit&&\omit&&\omit&&\omit&\cr
\noalign{\hrule}
&\omit&&\omit&&\omit&&\omit&&\omit&\cr

&$^{131}Xe$&&0.80&&20&&0.28&&80&\cr

&\omit&&\omit&&\omit&&\omit&&\omit&\cr}
\hrule}
$$

\vfill
\eject

{\bf 5. Neutralino Local Density}

  As for the value of the local neutralino density, $\rho_\chi$, to be
used in the rate of Eq.(2.2), for any point of the model parameter
space we take into account the relevant value of the neutralino
relic density. When $\Omega_\chi h^2$ is larger than a minimal
$(\Omega h^2)_{\rm min}$  consistent with the standard value for the local
dark matter density $\rho_l = {\rm 0.3~GeV~cm^{-3}}$, then we simply put
$\rho_\chi=\rho_l$. When $\Omega_\chi h^2$ turns out to be less than
$(\Omega h^2)_{\rm min}$,
we take

$$\rho_\chi = \rho_l {\Omega_\chi h^2 \over (\Omega h^2)_{\rm min}}
\equiv \rho_l \xi~. \eqno(5.1)$$

Here $(\Omega h^2)_{\rm min}$ is set equal to 0.05.
The effect of this
procedure of rescaling the local dark matter density according to the
actual neutralino relic abundance is manifest in some characteristic
features of our results, as is shown in the next section.

\bigskip
{\bf 6. Results }

{\bf 6.1 Germanium.}
We start the presentation of our results by discussing the event
rates expected for $Ge$--detectors. In Fig.1 we report the integrated
rate
$R( 2~\kev < \eee < 4~\kev) = \int_{2~\kev}^{4~\kev} d\eee dR/d\eee$
at the representative
point: $\tan \beta = 20$, $m_h = 50~\gev$,
$\msf = 1.2~m_\chi$, in the form of a
scatter plot obtained by varying the $M_2$, $\mu$ parameters in the ranges:
$20~\gev \leq M_2 \leq 6~\tev$
and $20~\gev \leq |\mu| \leq 3~\tev$. The range of the
neutralino mass $m_\chi$ considered here is
$20~\gev \leq m_\chi \leq 1~\tev$;
however, to simplify the presentation of our results, in Fig.1 and in
the following plots for $R$, only the range
$20~\gev \leq m_\chi \leq 300~\gev$ is
shown (at higher $m_\chi$ the plots retain  the same smooth decrease as
$m_\chi$ increases).

     To illustrate some of the features of the scatter plot in Fig.1,
let us recall two main points: i) at fixed $m_\chi$ the event rate $R$
depends on the model parameters only through the product
$\rho_\chi \sigma_i$ (see Eqs. (2.1)--(2.2)),
ii) this product is given by

$$\rho_\chi \sigma_i = \rho_l \sigma_i
\qquad {\rm when} \qquad
\Omega_\c h^2 \geq (\Omega h^2)_{\rm min} \eqno(6.1)$$

\noindent
and

$$\rho_\chi \sigma_i = \xi \rho_l \sigma_i
\propto (\den) \sigma_i
\propto {\sigma_i \over{< \sigma_{\rm ann} {\rm v} >}}
\qquad  {\rm when} \qquad
\den < (\Omega h^2)_{\rm min} \eqno(6.2)$$

\noindent
($\sigma_{\rm ann}$ and v being the
$\chi$--$\chi$ annihilation cross section and the
relative velocity; see eq.(3.1) of paper I).

In Fig.1 the wide spread in the values of $R$ at fixed $m_\chi$ is due to
the very large variations of
$\rho_\chi \sigma_i$ as we scan the $M_2$, $\mu$
plane along an isomass curve. A detailed analysis of our numerical
results shows that the maximal value of $R$, at each $m_\chi$, usually
corresponds to a point in the $M_2,\mu$ plane where $\Omega_\chi h^2$ is
substantially below $(\Omega h^2)_{\rm min}$. In other words, the maximal
expected signal is obtained for neutralino configurations which imply
use of Eq.(6.2) for $\rho_\chi$, with values for the rescaling parameter
$\xi$ which may also be very small. The nature of the neutralino
composition that at fixed $m_\chi$ provides the maximum $R$ depends on a very
crucial balance between $\sigma_i$ and $\sigma_{\rm ann}$
in the expression (6.2).
Indicatively, the largest event rates in the plot of Fig.1 entail the
following values for $\xi$:
$\xi \sim 0.01-0.02$ for $m_\chi \simeq 40-60~\gev$ (in
this case the gaugino--higgsino mixing is substantial), $\xi \sim 0.1$ for
$m_\chi \simeq 60-70~\gev$ (here the composition is mainly of the higgsino
type),
$\xi \sim 0.05$ for $m_\chi \simeq 70-90~\gev$
(large gaugino--higgsino mixing); above
$m_\chi \simeq 90~\gev$ the neutralino composition is mainly of the gaugino
type
with $\xi$ which reaches values of order of a few tenths.
These properties are depicted in Figs.2--3. Fig.2 shows
$\Omega_\chi h^2$ as
a function of $m_\chi$; the diamonds denote the values of the relic
abundances for those configurations that provide the maximal values of
$R$. The compositions of these neutralino configurations are shown in
the $M_2$--$\mu$ plane of Fig.3.

In view of these results we wish to emphasize that, by disregarding the
neutralino compositions that give low values of $\den$
(say, below $\Omega h^2 \sim 0.05$)
on the base that these compositions are not cosmologically interesting,
one would miss configurations that potentially generate the most sizable
signals in the direct searches.

It is worth noticing
that the configurations of maximal gaugino--higgsino mixing (defined in
paper I) provide large values of $R$ (denoted by circles in Fig.1) over
the whole $m_\chi$ range, even if the rescaling effect is always
substantial for these compositions; in fact in this case the elastic
$\chi$--nucleus coherent cross sections are very large and compensate for the
large values of $\sigma_{\rm ann}$ in Eq.(6.2).

  Because of the properties that we have just discussed, the graph of the
event rate $R$ (Fig.1) exhibits some of the salient
features of the plot of the relic abundance $\Omega_\chi h^2$ as a function
of $m_\chi$ (see Fig.2). In particular we recognize the pronounced
dips at $m_\chi \sim 25~\gev$ and at $m_\chi \sim 45~\gev$, i.e. at the values
where $\sigma_{\rm ann}$ has an $A(h)$--pole and a $Z$--pole, respectively.

    Finally we wish to stress that in the representative point
under discussion the maximal expected signal: $R \sim 6$ events/(Kg day),
which occurs at $m_\chi \sim 40~\gev$, differs only by a factor 1.5 from the
present experimental sensitivity $R_{\rm exp} \simeq 10$ events/(Kg day) [14]
(denoted by a horizontal
line in our plot of Fig.1). In order to illustrate the discovery
potential of the
direct neutralino search by $Ge$ detectors we show in Fig.4 the regions
in the $M_2$, $\mu$ plane that could be experimentally investigated in case
of improvements in sensitivity by a factor of 10
(heavy--dotted regions) or by a factor 100 (light--dotted regions).

In Fig.5 we give a sample of the differential rates $dR/d\eee$
corresponding to the neutralino compositions that, at a given $m_\chi$,
provide a maximal value for the integrated rate R, previously defined.
The values $m_\c = 40,~ 80,~ 150$ GeV are displayed in the figure together
with the present experimental points [14].

In Fig.6 we give the event rate $R$ at the same values of $\tan \beta$ and
$m_h$ as above, but for the second set of values for the sfermion masses, i.e.
$\msf = 3 ~ \tev$ . We notice that in this case the expected maximal rate is
even larger than in the previous example for $m_\chi > 70~ \gev$. This
increase is due to compositions which are more markedly of the gaugino
type (rescaling effect is small here, due to the suppression induced by
the high values of $\msf$  on the annihilation cross section). The relevant
explorable regions in the $M_2-\mu$ plane are shown in Fig.7.

Now let us see how the theoretical rates vary as we change the values
for $\tan \beta$ and for $m_h$. In our evaluations the Higgs--exchange graphs
have a substantial role; then, increasing $m_h$ has a suppression effect
on cross sections, and in particular on the elastic one. This is mainly
due to the dependence on the Higgs masses through the propagators
($\sim 1/m_h^4$).
$\tb$ enters in some of the coupling constants; it generally
occurs that cross sections decrease as $\tb$ decreases.
To illustrate the size of these effects, we show in Fig.8--9 the
integrated event rates for $\tb= 8$, $m_h = 50~ \gev$ for the two sets of
values for $\msf$ and in Fig.s 10--11 the relevant explorable regions in
the $M_2, \mu$ plane.
In Fig.s 12--13 $R$ is displayed at the representative point $\tb=8$,
$m_h=80 ~\gev$ for $\msf = 1.2~m_\chi$ and $\msf= 3~\tev$ respectively.
As compared
to the previous case $\tb = 8$, $m_h = 50~ \GeV$, the present event rate
$R$ is sizably lower, mainly because of the propagator effect
$\sim 1/m_h^4$. The regions in the $M_2, \mu$ plane which are experimentally
explorable, by increasing the sensitivity appropriately, are shown in
Fig.s 14--15.

We do not report here the results for other representative points.
We only quote that for instance at $\tb = 2$, as is expected, the explorable
regions are smaller than the previous ones. However,
a 2--order--of--magnitude improvement in sensitivity would still make
feasible an exploration in the range $20 ~\GeV \lsim \m \lsim 100 ~\GeV$.

Then, as far as the $Ge$ detectors are concerned, we can conclude that
improvements of about 2 orders of magnitude in experimental sensitivities
would allow investigation of the neutralino dark matter in large regions
of the model parameter space. It is extremely encouraging that
this level of sensitivities appears to be within reach in the
near future [20].

Let us turn now to a brief discussion of the other materials mentioned
before. The theoretical analysis has been performed along the same lines
as in the case of $Ge$. For sake of brevity, here we only report the
differential and the integrated rates for the two representative points:
$\tb = 20$, $m_h = 50 ~\GeV$, $\msf = 1.2~m_\chi$ and $\tb = 20$,
$m_h = 50 ~\GeV$, $\msf = 3 ~\TeV$.

{\bf 6.2 Sodium Iodide}.
Theoretical expectations for this diatomic material were already
presented and compared with experimental values in Ref.[2] for
neutralino masses in the range $20 ~\GeV \leq m_\chi \leq 80 ~\GeV$. Here the
analysis is extended up to $m_\chi = 1 ~\TeV$ and to much wider regions of
the parameter space. The results of the integrated rate
$R(4.5 ~\KeV \leq \eee \leq 5 ~\KeV)$ are shown in Fig.s 16--17.

For the case of $\tb = 20$, $m_h = 50 ~\GeV$,
$\msf = 1.2~m_\chi$ we also give in Fig. 18 the maximal differential rate
at three values of $m_\chi$. We notice that the peculiar behaviour of
$dR/d\eee$ as a function of $m_\chi$ is due to the diatomic nature of $NaI$.
The response of a $NaI$ detector to the {\n} scattering is dominated by
Iodine
up to $\eee \sim 6 ~\KeV$ and by Sodium at higher $\eee$.
This circumstance depends
on the very different coherent form factor effects in the two nuclei as
is shown by the values reported in Table I. From the threshold up to
$\eee \sim 6 ~\KeV$ in Iodine the large suppression due to
$F^2(E_R^{\rm th})$
is compensated by a large value of its coherent pointlike cross section
($A^2$ effect). At higher values of $\eee$ the Iodine coherent form factor
has a fast fall--off, because of the zero at $\eee = 7.4~\KeV$, whereas in
the whole range of $\eee$ shown in Fig. 18 the Sodium form factor has a rather
flat behaviour.

As discussed in
Ref.[2] an improvement of two orders of magnitude in the $NaI$ detectors
sensitivity is foreseeable in the next few years. This would allow to
start the exploration of the physics of dark matter neutralinos with
these detectors.

{\bf 6.3 Xenon}.
Much experimental activity is now being developed in dark matter
detectors which employ $Xe$. We report in
Fig. 19--20 the integrated rate
$R(20 ~\KeV \leq \eee \leq 22 ~\KeV)$ in the representative points:
$\tb = 20$, $m_h = 50 ~\GeV$, $\msf = 1.2~m_\chi$ and $\tb = 20$,
$m_h = 50 ~\GeV$, $\msf = 3 ~\TeV$. For the first of these
points also the differential rate is given (Fig. 21). We notice that
for $Xe$ the suppression due to the coherent form factor is sizeable
(see Table 1). However the use of $Xe$ for detecting dark matter
neutralinos appear to be very promising in view of the experimental
features of these detectors.

{\bf 6.4 Calcium Fluoride}.
Finally in Fig.s 22--23 we display the integrated rate
$R(8.5 ~\KeV \leq \eee \leq 9 ~\KeV)$ for $CaF_2$ in the usual
representative points:
$\tb = 20$, $m_h = 50 ~\GeV$, $\msf = 1.2~m_\chi$ and $\tb = 20$,
$m_h = 50 ~\GeV$, $\msf = 3 ~\TeV$.  In Fig. 24 also the
differential rate for the first representative point is shown.

{\bf 7. Conclusions}

We can conclude that the use of different materials and the achievements
of improved experimental performances are offering very exciting perspectives
for a direct search of {\n} as a dark matter component.

The theoretical evaluations presented here aim at setting the level of the
experimental sensitivities required to start the exploration of the
physics of dark matter \n s. The quantitative relations between the
experimental sensitivity of a particular detector and its discovery potential
are easily readable in our plots.

Some final comments on the main theoretical uncertainties
are in order here: \hfill \break
i) major improvements in theoretical predictive power require some experimental
informations about the masses of particles to be discovered at accelerators:
Higgs bosons, SuSy particles; \hfill \break
ii) the evaluation of the elastic coherent cross section suffer from an
uncertainty in the estimate of the Higgs--nucleon strength. In Ref.[3] we
have analyzed the relevance of this effect for the direct {\n} search, by
comparing different estimates for the Higgs--nucleon strength [21,22].
A new recent study of this fundamental coupling has shown that the problem
needs further investigation [23]. In the present paper we have employed
for the Higgs--nucleon coupling the estimate of Ref.[21]. Use of the estimate
in Ref.[22] would reduce our rates
by a factor of 3 approximately; \hfill \break
iii) the value of $(\Omega h^2)_{\rm min}$ to be employed for rescaling the
local {\n} density is obviously somewhat arbitrary. In the present paper we
have conservatively taken $(\Omega h^2)_{\rm min}=0.05$. In view of the small
value of $\Omega$ which appears to be implied by the estimated amount
of dark matter in our galaxy, our values for $(\Omega h^2)_{\rm min}$ could
quite well be reduced by a factor of 2. This reduction in
$(\Omega h^2)_{\rm min}$ would increase the maximal rate $R$ by a factor of 2,
since, as we have seen above, maximal signals occur in general for {\n}
compositions where the rescaling effect is large.

\bigskip
\centerline{*\ \ \  *\ \ \  *}
\bigskip

This work was supported in part by Research Funds of the Ministero
dell'Universit\`a e della Ricerca Scientifica e Tecnologica.

\vfill
\eject

\centerline{\bf References}

\item{[1]}
A.Bottino, V.de Alfaro, N.Fornengo,
G.Mignola and M.Pignone,
On the Neutralino as Dark Matter Candidate. I. Relic Abundance,
University of Torino preprint DFTT 37/93.

\item{[2]}
A.Bottino, V.de Alfaro, N.Fornengo, G.Mignola, S.Scopel, and
C.Bacci et al. (BRS Collaboration), Phys. Lett. B295(1992)330.

\item{[3]}
A.Bottino, V.de Alfaro, N.Fornengo, A.Morales, J.Puimedon and
S.Scopel, Modern Physics Letters A7(1992)733.

\item{[4]}
R. Barbieri, M. Frigeni and G.F. Giudice,
Nucl. Phys. B313(1989)725.

\item{[5]}
K.Griest, Phys. Rev. D38(1988)2357;
Phys. Rev. Lett. 61(1988)666.

\item{[6]}
L. G. Pondrom (CDF Collaboration), presented at XXV Int. Conf. on
High Energy Physics, Singapore 1990, published in Singapore HEP (1990)144.

\item{[7]}
F.Abe et al. (CDF Collaboration), Phys. Rev. Lett. 69(1992)3439.

\item{[8]}
H.Baer, X.Tata and J.Woodside, Phys. Rev. D44(1991)207.

\item{[9]}
D.Decamp et al. (ALEPH Coll.), Phys. Rep. 216(1992)253.

\item{[10]}
J.Ellis and R.Flores,
Phys. Lett. B263 (1991) 259;
Phys. Lett. B300 (1993) 175.

\item{[11]}
J.Engel and P.Vogel, Phys. Rev. D40 (1989) 3132 \hfill \break
A.F.Pacheco and D.D.Strottman, Phys.Rev.D40 (1989) 2131; \hfill \break
F.Iachello, L.M.Krauss and G.Maino, Phys. Lett. B254 (1991) 220.

\item{[12]}
J.Engel, Phys. Lett. B264 (1991) 114.

\item{[13]}
The values of the quenching factors $Q$ and of the
threshold energies $E_R^{\rm th}$ quoted in Table 1 are
taken from: $Ge$ [14--15],
$NaI$ [16,2], $Xe$ [17], $CaF_2$ (private communications by
P.Belli, R.Bernabei, G.Gerbier).

\item{[14]}
D.Reusser et al., Phys. Lett. B255 (1991) 143

\item{[15]}
D.O. Caldwell et al., Phys. Rev. Lett., 61 (1988) 510; \hfill \break
S.P. Ahlen et al., Phys. Lett. B195 (1987) 603; \hfill \break
E. Garc\'\i a et al., Nucl. Phys. B (Proc. Suppl.) 28A (1992) 286

\item{[16]}
P. Belli et al., Phys. Lett. B293 (1992) 460

\item{[17]}
P. Belli et al., LNGS -- 93/55 May 1993 Preprint; \hfill \break
P. Belli et al., Nucl. Instr. Meth. Phys. Res. A316 (1992) 55.

\item{[18]}
J.Engel, S.Pittel, E.Ormand and P.Vogel, Phys Lett. B275
(1992) 119.

\item{[19]}
M.T. Ressel, M.B. Auferheide, S.D. Bloom, K. Griest, G.J. Mathews, D.A. Resler,
preprint UCRL--JC--114085, 1993.

\item {[20]}
For an up--to--date review of the present achievements and the new
perspectives for dark matter detectors using $Ge$ as well as the other
materials discussed in the present work, see the experimental papers in
the Proceedings of the Workshop on
{\it ``The Dark Side of the Universe''}, Rome, June 1993,
(World Scientific Publishing Co.,to appear).

\item {[21]}
T.P. Cheng, Phys. Rev D38 (1988) 2869;
H.-Y. Cheng, Phys. Lett. B219 (1989) 347.

\item{[22]}
J.Gasser, H.Leutwyler and M.E.Sainio,
Phys. Lett. B253 (1991) 252.

\item{[23]}
M.Drees, M.M.Nojiri, Phys. Rev. D47(1993) 4226.

\vfill
\eject
\centerline{\bf Table Caption}
\bigskip
{\bf Table 1}.
Quenching factors $Q$, threshold energies $\eee^{\rm th}$, squared values of
the form factor (4.1) at the threshold $F^2(E_R^{\rm th})$ and the
values $\eee^0$ where the first zero of Eq.(4.1) occurs
are listed for the nuclei considered
in our analysis.
\vfill
\eject
\centerline{\bf Figure Captions}

\bigskip

{\bf Figure 1}.
Scatter plot of the integrated rate $R(2~\kev \leq \eee \leq 4~\KeV)$
for neutralino--$Ge$
interaction as a function
of neutralino mass $\m$, at the representative point:
$\tb=20$, $\mh=50~\gev$ and $\msf=1.2~\m$.
The parameters $M_2$ and $\mu$ are varied in
the ranges ${\rm 20~GeV} \leq M_2 \leq {\rm 6~TeV}$ and
${\rm 20~GeV} \leq  |\mu| \leq {\rm 3~TeV}$.
The horizontal line denotes the present experimental limit $R_{\rm exp}$.
Filled--circles represent the $R$ values  for  neutralino
compositions where gaugino and higgsino components are maximally mixed
($0.45 \leq a_1^2+a_2^2 \leq 0.55$).

\bigskip

{\bf Figure 2}.
Scatter plot of the neutralino relic abundance $\den$ as a function
of neutralino mass $\m$. The relic abundance is calculated in the
same representative point as in
Fig.1 and also with the same variations over $M_2,\mu$.
The horizontal line denotes the value
$(\Omega h^2)_{\rm min}=0.05$.
Diamonds represent the values of relic abundance which give the
maximal value of $R$ for a given neutralino mass $m_\chi$
(they correspond to the maxima of $R$ in Fig.1).

\bigskip

{\bf Figure 3}.
Isomass curves and composition lines for neutralino in the
$M_2-\mu$ plane for $\tan \beta = 20$.
Dashed lines are
lines of constant neutralino mass ($m_\chi$ = 30 \ GeV, 100 \ GeV,
300 \ GeV and 1 \ TeV). Solid lines refer to constant gaugino
fraction $f_g$ in the neutralino composition
($f_g = a_1^2 +a_2^2$): $f_g$ = 0.99, 0.9,
0.5, 0.1 and 0.01.
Squares represent the points in the $M_2-\mu$ parameter space which give the
maximal integrated rate $R$ for a given neutralino mass $m_\chi$
(they correspond to the maxima of $R$ in Fig. 1).

\bigskip

{\bf Figure 4}.
Explorable regions in the $M_2-\mu$ parameter space for
neutralino--$Ge$ interaction. Parameters are:
${\rm 20~GeV} \leq  M_2 \leq  {\rm 6~TeV}$, ${\rm 20~GeV} \leq |\mu| \leq
{\rm 3~TeV}$,
$\tb=20$, $\mh=50~\gev$ and $\msf=1.2~\m$.
Neutralino masses extend up to $\m=1~\tev$.
Heavy--dots denote regions where the integrated rate $R$
is in the range $0.1~R_{\rm exp} < R \leq R_{\rm exp}$, being $R_{\rm exp}$
the present experimental bound. In light--dotted regions
$0.01~R_{\rm exp} \leq R \leq 0.1~R_{\rm exp}$ occurs.
Lines denote curves of constant neutralino mass: dashed lines stand for
$m_\chi=20~\gev$, solid lines for $m_\chi=100~\gev$,
dot--dashed lines for $m_\chi=300~\gev$ and
dotted lines for $m_\chi=1~\tev$.

\bigskip

{\bf Figure 5}.
Nuclear recoil spectrum $dR/d\eee$ for neutralino--$Ge$ interaction
as a function of the electron--equivalent energy $\eee$.
Lines denote the nuclear recoil spectra which give the maximal integrated rate
$R$ for a fixed neutralino mass: solid line is calculated for $\m=40~\gev$,
dashed line for $\m=80~\gev$ and dot--dashed line for $\m=150~\gev$.
$M_2$ and $\mu$ parameters are varied in the ranges
${\rm 20~GeV} \leq M_2 \leq {\rm 6~TeV}$ and
${\rm 20~GeV} \leq |\mu| \leq {\rm 3~TeV}$.
The other parameters are: $\tb=20$, $\mh=50~\gev$ and $\msf=1.2~\m$.
Crosses represent the present experimental limit [14].

\bigskip

{\bf Figure 6}.
Same as in Fig. 1, except for $\msf=3~\tev$ instead of $\msf=1.2~\m$.

\bigskip

{\bf Figure 7}.
Same as in Fig. 4, except for $\msf=3~\tev$ instead of $\msf=1.2~\m$.

\bigskip

{\bf Figure 8}.
Same as in Fig.1; here the representative point is:
$\tb=8$, $\mh=50~\gev$ and $\msf=1.2~\m$.

\bigskip

{\bf Figure 9}.
Same as in Fig.1, here the representative point is:
$\tb=8$, $\mh=50~\gev$ and $\msf=3~\tev$.

\bigskip

{\bf Figure 10}.
Same as in Fig.4, here the representative point is:
$\tb=8$, $\mh=50~\gev$ and $\msf=1.2~\m$.

\bigskip

{\bf Figure 11}.
Same as in Fig.4, here the representative point is:
$\tb=8$, $\mh=50~\gev$ and $\msf=3~\tev$.

\bigskip

{\bf Figure 12}.
Same as in Fig.1, here the representative point is:
$\tb=8$, $\mh=80~\gev$ and $\msf=1.2~\m$.

\bigskip

{\bf Figure 13}.
Same as in Fig.1, here the representative point is:
$\tb=8$, $\mh=80~\gev$ and $\msf=3~\tev$.

\bigskip

{\bf Figure 14}.
Same as in Fig.4, here the representative point is:
$\tb=8$, $\mh=80~\gev$ and $\msf=1.2~\m$.

\bigskip

{\bf Figure 15}.
Same as in Fig.4, here the representative point is:
$\tb=8$, $\mh=80~\gev$ and $\msf=3~\tev$.

\bigskip

{\bf Figure 16}.
Scatter plot of the integrated rate $R(4.5~\kev \leq \eee \leq 5~\KeV)$
for neutralino--$NaI$ interaction.
The representative point is:
$\tb=20$, $\mh=50~\gev$ and $\msf=1.2~\m$;
the parameters $M_2$ and $\mu$ are varied in
the ranges ${\rm 20~GeV} \leq M_2 \leq {\rm 6~TeV}$ and
${\rm 20~GeV} \leq  |\mu| \leq {\rm 3~TeV}$.
Filled--circles represent the $R$ values  for  neutralino
compositions where gaugino and higgsino components are maximally mixed
($0.45 \leq a_1^2+a_2^2 \leq 0.55$).

\bigskip

{\bf Figure 17}.
Same as in Fig. 16, except for $\msf=3~\tev$ instead of $\msf=1.2~\m$.

\bigskip

{\bf Figure 18}.
Nuclear recoil spectrum $dR/d\eee$ for neutralino--$NaI$ interaction
as a function of the electron--equivalent energy $\eee$.
Lines denote the nuclear recoil spectra which give the maximal
integrated rate
for a fixed neutralino mass: solid line is calculated for $\m=40~\gev$,
dashed line for $\m=80~\gev$ and dot--dashed line for $\m=150~\gev$.
$M_2$ and $\mu$ parameters are varied in the ranges
${\rm 20~GeV} \leq M_2 \leq {\rm 6~TeV}$ and
${\rm 20~GeV} \leq |\mu| \leq {\rm 3~TeV}$.
The other parameters are: $\tb=20$, $\mh=50~\gev$ and $\msf=1.2~\m$.
Crosses represent the present experimental limit [16].

\bigskip

{\bf Figure 19}.
Scatter plot of the integrated rate $R(20~\kev \leq \eee \leq 22~\KeV)$
for neutralino--$Xe$ interaction.
The values for the model parameters and the notations are as in Fig.16.

\bigskip

{\bf Figure 20}.
Same as in Fig. 19, except for $\msf=3~\tev$ instead of $\msf=1.2~\m$.

\bigskip

{\bf Figure 21}.
Nuclear recoil spectrum $dR/d\eee$ for neutralino--$Xe$ interaction
as a function of the electron--equivalent energy $\eee$.
The values for the model parameters and the notations are as in Fig.18.
Here no experimental data are available.

\bigskip

{\bf Figure 22}.
Scatter plot of the integrated rate $R(8.5~\kev \leq \eee \leq 9~\KeV)$
for neutralino--$CaF_2$ interaction.
The values for the model parameters and the notations are as in Fig.16.

\bigskip

{\bf Figure 23}.
Same as in Fig. 22, except for $\msf=3~\tev$ instead of $\msf=1.2~\m$.

\bigskip

{\bf Figure 24}.
Nuclear recoil spectrum $dR/d\eee$ for neutralino--$CaF_2$ interaction
as a function of the electron--equivalent energy $\eee$.
The values for the model parameters and the notations are as in Fig.18.
Here no experimental data are available.

\bye